\documentclass[prd,nofootinbib,noshowkeys,superscriptaddress]{revtex4}
\usepackage[latin1]{inputenc}
\usepackage{pstricks,pst-node,pst-text,pst-3d,pslatex}
\usepackage[T1]{fontenc}

\usepackage{graphicx}
\usepackage{amsfonts}
\usepackage{amsmath}

\newcommand{\be}{\begin{eqnarray}}
\newcommand{\ee}{\end{eqnarray}}

 \newcommand{\x}{\tilde X}
 \newcommand{\om}{\omega}
\newcommand{\Om}{\Omega}

\begin{document}

\title{Molecular Dynamics at Low Time Resolution} 

\author{P. Faccioli}
\affiliation{Dipartimento di Fisica  Universit\'a degli Studi di Trento, Via Sommarive 14, Povo (Trento), I-38050 Italy.}
\affiliation{INFN, Gruppo Collegato di Trento, Via Sommarive 14, Povo (Trento), I-38050 Italy.} 
\email{faccioli@science.unitn.it}

\begin{abstract}
The internal dynamics of macro-molecular systems is characterized by widely separated time scales, ranging from fraction of ps to ns.  
In ordinary molecular dynamics simulations, the elementary time step $\Delta t$ used to integrate the equation of motion needs to be  chosen much smaller of the shortest time scale, in order not to cut-off important physical effects.
We show that, in systems obeying the over-damped Langevin Eq., the fast molecular dynamics which occurs at time scales smaller than  $\Delta t$ can be analytically integrated out and gives raise to a time-dependent correction to the diffusion coefficient, which we rigorously compute.
The resulting effective Langevin equation describes by construction the same long-time dynamics, but has a lower time resolution power,  hence it can be integrated using larger time steps $\Delta t$.  
We illustrate and validate this method by studying the diffusion of a point-particle in a one-dimensional toy-model and the  denaturation of a protein.
\end{abstract}

\maketitle

\section{Introduction}

Molecular Dynamics (MD) simulations are playing an increasingly important role in contemporary biophysics,  biochemistry and molecular biology, as they allow  for an atomistic level of description of many fundamental molecular processes. 
Unfortunately, when the  system  is very large, or when the reaction under investigation is very slow,  the computational cost of the MD simulations can be extremely large.
 Hence,  a large effort is being invested by several groups,  in order to develop alternative theoretical approaches or improved numerical algorithms. 
  Examples include reaction path sampling methods \cite{TPS, Elber, DRP, elber2, Donniach}, Markov state models \cite{MSM1,MSM2,MSM3}, projection techniques \cite{projection}, adaptive time-step MD \cite{adaptive} and temperature accelerated MD \cite{TMD}, to name a few. 

The inefficiency of MD simulations to investigate the long-time dynamics is related to the co-existence of  widely separated time scales in molecular systems.  For example, while the time scale for equilibrium oscillations of covalent bonds is of the order of the fraction of ps, the time scale associated to the rotation of dihedral angles in a poly-peptide chain is of the order of ns. Clearly, in order to perform realistic MD simulations, one has to use integration time steps which are much shorter than the shortest time scale,  hence typically in the fs range. 

In a recent work \cite{RG1} it was shown that such a separation of time scales can be exploited in order to rigorously derive an effective stochastic theory (EST)  which generates by construction the same long-time dynamics of  the ordinary Langevin equation (LE), at a lower time resolution power. 
The basic idea of the EST consists in using the path integral formalism and  Renormalization Group (RG) techniques, in order to systematically and analytically perform the integral over the fast Fourier components of the Langevin trajectories.  
The advantage of this procedure is that the low time resolution EST can be simulated using larger discretization time steps. It is important to emphasize that the EST is formulated in terms of the same degrees of freedom of the original theory (e.g. the atomic coordinates), hence not rely on any choice of reaction coordinate.

In  \cite{RG1} the EST was formulated in terms of a stochastic path integral and tested on a simple one-dimensional toy model, using a Monte Carlo algorithm. 
The main result of the present work is to formulate such an effective theory in terms of an effective Langevin equation, which contains a time-dependent diffusion constant.  Such an equation can  be straightforwardly integrated using a standard Ito rule, and adopting large time steps $\Delta t$. 

In order to illustrate and validate the present approach, we apply it  to simulate the stochastic dynamics of two test-systems: a simple one-dimensional toy model and a small protein fragment. In both cases we find that the effective Langevin equation yields the correct long-time evolution of the system, even when one adopts an integration time step 20 times larger than that used to integrate the original LE. 

The paper is organized as follows. In section \ref{sto} we review the path integral formulation of the ordinary over-damped Langevin dynamics. In section \ref{ESTpi} we summarize the construction of the path integral which defines the EST.  In section \ref{ESTmd}, we show how such an effective theory can be equivalently formulated in terms of an effective Langevin equation, in which the physical effects associated to the fast dynamics is replaced by an effective diffusion coefficient. 
In section \ref{illustrative} we present  the  illustrative applications of the present approach. Our conclusions are summarized in section \ref{conclusions}.

\section{Path Integral Formulation of the Over-Damped Langevin Dynamics}
\label{sto}

Let us  begin by reviewing the path integral formulation of the stochastic dynamics generated by the ordinary over-damped Langevin~Eq.  
\be
\dot {\bf X} = - \frac{D_0}{k_B T} \nabla U({\bf X}) + {\bf \eta}(t).
\label{lang}
\ee 
Here ${\bf X}=({\bf x_1},\ldots, {\bf x}_{N_p})$ is a point in the  configuration space  of $N_p$ particles, $D_0$ is the diffusion constant and ${\bf \eta}(t)$ is a stochastic force with zero average, obeying fluctuation-dissipation relationship 
\be
\langle \eta^a_i(t) \eta^b_j(0)\rangle = 2 D_0~ \delta(t) ~\delta_{i j}~\delta^{a b},\qquad~(i,j=1, \ldots, N_p~ \textrm{~} ~ a,b=1,2,3).
\ee

The stochastic differential Eq. (\ref{lang}) needs to be complemented by a prescription which provides a definition of the time derivative of the stochastic variable ${\bf X}$. 
A choice which is commonly adopted in numerical simulations is  the so-called \emph{Ito prescription}, in which Eq. (\ref{lang}) is defined as the limit of the discretized equation
\be
{\bf X}_{l+1} - {\bf X}_l =  - \frac{D_0~\Delta t}{k_B T} \nabla U({\bf X}_l) + \sqrt{2 D_0 \Delta t}~ {\bf R}_l,
\label{ito}
\ee
where ${\bf X}_l$ represents the configuration at the $l-$th time step and ${\bf R}_l$ is a Gaussian noise vector with zero average and variance given by
\be
\label{RR}
\langle  R^a_{i~l} ~R^b_{j~l'} \rangle = ~\delta_{i j} ~\delta_{l l'}~\delta^{a b}.
\ee

We note that the Eq. (\ref{ito}) defines a Markovian process, i.e. the probability distribution for the configuration at time step $l+1$ is completely determined by the configuration at  time step $l$. The term $\sqrt{2 D_0 \Delta t}~ {\bf R}_i$ is the random (Brownian) displacement in configuration space, after an elementary time interval $\Delta t$.

Let us now compute the probability a given path, i.e. of a specific sequence of $N_t$ 
conformations ${\bf X}(\tau) \equiv ({\bf X}_1, {\bf X}_2, \ldots, {\bf X}_{N_t})$, generated by iterating $N_t$ times Eq. (\ref{ito}). To this end, we follow closely the discussion in \cite{elber2} and  first  compute 
the (normalized) probability density of generating a given string of random numbers $({\bf R}_1, {\bf R}_2, \ldots, {\bf R}_{N_t})$ 
\be
\mathcal{P}({\bf R}_1, \ldots, {\bf R}_{N_t}) = \left(\frac{1}{2 \pi}\right)^{\frac{3  N_p N_t}{2}}~\prod_{j=1}^{N_t}~\exp\left[-\frac{{\bf R}^2_j}{2}\right].
~\prod_{i=1}^{N_t} d {\bf R}_i
\label{p1}
\ee
Next, we use Eq. (\ref{ito}) to relate the probability density of such a sequence of random numbers  to the probability density of the sequence of 
configurations $({\bf X}_1, {\bf X}_2, \ldots, {\bf X}_{N_t})$. After substituting ({\ref{ito}) into the exponents and computing the Jacobian of such a transformation, one arrives to:
\be
\mathcal{P}( {\bf X}_1,\ldots, {\bf X}_{N_p}) = \text{const.}~\times~ e^{-\frac{-\sum_{l=1}^{N_t-1}
\left({\bf X}_{l+1} - {\bf X}_{l} + D_0~\frac{\Delta t}{k_B T} \nabla U({\bf X}_l)\right)^2}{4 D_0 \Delta t}}~\prod_{i=1}^{N_t} d{\bf X}_i.
\label{pathP}
\ee

The (un-normalized) conditional probability of visiting the configuration ${\bf X}_{N_t}$ after $N_t$ steps, starting from the configuration ${\bf X}_1$ reads:
\be
P( {\bf X}_1| {\bf X}_{N_p}; N_t \Delta t) = \int ~\left(\prod_{i=1}^{N_t} d{\bf X}_i\right) ~ e^{-\frac{-\sum_{l=1}^{N_t-1}
\left({\bf X}_{l+1} - {\bf X}_{l} + D_0~\frac{\Delta t}{k_B T} \nabla U({\bf X}_l)\right)^2}{4 D_0 \Delta t}}
\label{pathing}
\ee

In the continuum limit, 
\be
N_t~&\to& \infty,\\
N_t \Delta t &\equiv& t \qquad(\textrm{fixed}),
\ee
 the   expansion of the square in the exponent of Eq. (\ref{pathP}) contains the definition of the integral in the  so-called Ito stochastic Calculus:
\be
(I) \int_{{\bf X}_1}^{{\bf X}_N} \nabla U({\bf X})\cdot d {\bf X} \equiv \lim_{\stackrel{N\to \infty}{ \Delta t\to0}} \sum_{l=1}^{N-1}\left({\bf X}_{l+1} - {\bf X}_{l}) \cdot \nabla U({\bf X}_l \right).
\ee
Due to the stochastic nature of the variable ${\bf X}$, the fundamental theorem of the Ito Calculus differs from the one of the Riemann Calculus, and reads ---see e.g. the 
discussion in  \cite{adib}---
\be
(I) \int_{{\bf X}_1}^{{\bf X}_N}~\nabla U({\bf X})\cdot   d {\bf X} = U({\bf X}_N) - U({\bf X}_1) -D_0~\int_{0}^{t} d\tau ~\nabla^2 U({\bf X}(\tau)),
\label{ftic}
\ee
where the integral appearing on the right-hand-side is an ordinary Riemann integral. Notice that, in the limit $D_0\to0$ limit, one recovers the conventional fundamental theorem of  Calculus.

Using this theorem, in the continuum limit  the conditional probability $P({\bf X}_N|{\bf X}_1; t= N\Delta t) $ can be expressed as a path integral
\be
P({\bf X}_N|{\bf X}_1; t)  = e^{-\frac{1}{2 k_B t} \left(U({\bf X}_N) - U({\bf X}_1)\right)} ~\int_{{\bf X}_1}^{{\bf X}_N} \mathcal{D} {\bf X} ~e^{- S_{eff}[{\bf X}]}, 
\label{PI}
\ee
where  
\be
S_{eff}[{\bf X}(t)]=\int_0^t d\tau \left(~\frac{\dot {\bf X}^2}{4 D_0} + V_{eff}({\bf X})\right)
\ee
is called the effective action and 
\be
\label{Veff}
V_{eff}({\bf X}) =  \frac{D_0}{4 (k_B T)^2} \left((\nabla U({\bf X}))^2 - 2 k_B T \nabla^2 U({\bf X})\right)
\ee
is called the effective potential. 

Incidentally, we note that the same expression (\ref{PI}) can be obtained directly from the Fokker-Planck Eq., without having to define a stochastic Calculus ---see e.g. the discussion in \cite{DRP}---.  In general,  it can been shown that the probability density generated by the  Langevin Eq. with a  \emph{constant} diffusion coefficient $D_0$ is independent on the convention adopted to define the stochastic Calculus. As we shall see in section \ref{ESTmd}, this is not the case for stochastic differential equations,  with a multiplicative noise.

 \section{The effective path integral for the long-time stochastic dynamics} 
 \label{ESTpi}
 
 In this section, we sketch the derivation of the EST, which formulates the stochastic dynamics described by the path integral (\ref{PI}) at a lower time resolution power. For all further details we refer the reader to the original paper \cite{RG1}.

 For simplicity and without loss of generality, it is convenient to consider the path integral with periodic boundary conditions    
\be
Z(t) \equiv \int d {\bf X} ~P({\bf X}| {\bf X};t) =  \oint \mathcal{D} {\bf X} ~e^{- S_{eff}[{\bf X}]}.
\label{pPI}
\ee
Let us introduce the Fourier components of the paths, 
\be
{\bf \x}(\omega_n) &=& \frac{1}{t}~\int_0^t d\tau ~{\bf X}(\tau)~e^{- i \omega_n t }  \\
{\bf X}(\tau) &=& {\bf X}(\tau+t) =  \sum_n {\bf \x}(\om_n) ~ e^{ i \omega_n t }.
\ee
where $\om_n$ are the Fourier  frequencies:
\be
\omega_n = \frac{2 \pi}{t}~n, \qquad  n=0, \pm 1, \pm 2, \ldots.  
\ee

The path integral (\ref{pPI}) is defined in the continuum limit. In practice, numerical simulations are always performed using a finite discretization time step $\Delta t$. 
Clearly, the shortest time intervals which can be explored in a numerical simulation is of the order of few  $\Delta t$. Equivalently,  the largest frequencies of the  Fourier transform of the stochastic paths $\tilde {\bf X}(\omega)$ are of the order few fractions of an ultra-violet (UV) cut-off $$\Omega \equiv 2 \pi/ \Delta t.$$
 
In order to exploit the decoupling of the internal time scales in molecular systems,  it is convenient to split the Fourier modes of the paths contributing to (\ref{PI})  in high-frequency ---or \emph{"fast"}--- modes and low-frequency ---or \emph{"slow"}--- modes. In this case, a real number  $0<b<1$ can be defined  such that the 
frequency  range $(0, \Om)$ is split in two intervals $(0, b~\Omega)\cup~(b~\Om, \Om)$. Correspondingly, one can define the "fast" component of the path ${\bf X}_>(\tau)$ and the "slow" component of the path ${\bf X}_<(\tau)$, by summing over the Fourier modes in the  $(0, b~\Omega)$ and $(b~\Om, \Om)$ range, respectively: 
 \be
{\bf X}_<(t) &=& \sum_{|\om_n| \le b\Om}~{\bf \x}(\omega_n) ~e^{~i \om_n t }\\
 {\bf X}_>(t) &=& \sum_{b\Om\le |\om_n| \le \Om}~{\bf \x}(\omega_n) ~e^{~i \om_n t }.
  \ee

The complete path integral  (\ref{pPI}) can therefore be written in the following way:
\be
\label{effAct}
Z(t)~&=&  \oint   \mathcal{D}{\bf X}_<\oint \mathcal{D}{\bf X}_>~ e^{-S_{eff}[{\bf X}_<+{\bf X}_>]}\nonumber\\
&\equiv& \oint  \mathcal{D}{\bf X}_<~ e^{-S_{eff}[{\bf X}_<]}~e^{- S_>[{\bf X}_<]},
\ee
where 
\be
\label{path2}
e^{- S_>[{\bf X}_<(\tau)]}\equiv \oint \mathcal{D}{\bf X}_> e^{S_{eff}({\bf X}_<] - S_{eff}[{\bf X}_<+{\bf X}_>]}
\ee
is called the renormalized part of the effective action. 

The EST is constructed by explicitly evaluating $S_{>}[{\bf X}_<]$, i.e.  by performing the path integral over fast modes ${\bf X}_>(\tau)$. In the limit in which the fast and slow modes are separated by a large gap in the spectrum of Fourier modes --- i.e. if the system displays a decoupling of time scales---such an integral can be carried out analytically in a perturbative approach based on Feynmann diagram techniques\cite{RG1}. 
The expansion parameter such a perturbation theory is the ratio between the typical frequency $\omega$ of the slow modes  and the UV cut-off  $b\Omega$.
Clearly, if hard and slow modes are decoupled, the ratio $\omega/(b\Omega)$ is a small number, hence the terms proportional to higher and higher powers  $L$ of  such a ratio  provide smaller and smaller corrections. 

If one accounts only for corrections up to order $L=3$, the renormalized part of the action takes the form of an effective interaction term~\cite{RG1}, i.e. 
\be
e^{- S_>[{\bf X}_<(\tau)]} = e^{-\int_0^t d\tau ~V_{eff}^R[{\bf X}_<(\tau)]}
\ee
where 
\be
V^{R}_{eff} ({\bf X}) &\simeq& 
\frac{D_0~(1-b)}{~\pi~ b\Omega} ~\nabla^2 V_{eff}({\bf X})\nonumber\\
&+& \frac{1}{2 }\left(\frac{D_0 (1-b)}{~\pi~b\Omega}\right)^2 ~\nabla^4 V_{eff}({\bf X})\nonumber\\
&+& \frac{1}{6 }\left(\frac{D_0~(1-b)}{\pi b\Omega}\right)^3 ~\nabla^6 V_{eff}({\bf X})
-\frac{D_0^2(1-b^3)}{3\pi~(b\Omega)^3}
\Big(\partial_i\partial_jV_{eff}({\bf X}) \Big)^2.
\label{L-leq-3}
\ee   
Note that the first line is the leading order term (i.e. $ L=1$) , while the second and third lines display the order $L=2$ and $L=3$ corrections, respectively.

We emphasize that the result of the EST construction is a new expression for the \emph{same} path integral (\ref{pPI}), in which the UV cutoff been lowered from $\Omega$ to $b \Omega$. Equivalently, the  path integral is discretized according to a  larger elementary  time step,  $ \Delta t \to \Delta t/b$: 
\be
Z^{\Delta t}(t) &\equiv& \oint_{\Delta t} \mathcal{D} {\bf X} ~e^{- S_{eff}[{\bf X}]} \propto \oint_{\Delta t/b} \mathcal{D} {\bf X} ~e^{- S_{eff}[{\bf X}] - 
 \int_0^t d\tau ~V^R_{eff}[{\bf X}(\tau)]} \equiv Z^{\Delta t/b}_{EST}(t)
\label{EST}
\ee
In these expressions, the symbol $\oint_{\Delta t}$ denotes the fact that the path integral is discretized according to an elementary time step $\Delta t$ and we have suppressed the subscript "<",  in the paths. It can be shown that the proportionality factor between $Z^{\Delta t}(t)$ and $Z^{\Delta t/b}_{EST}(t)$ depends only on $t$ and does not
contribute to the statistical averages.

We conclude this section be emphasizing that the EST is expected to accurately describe only the long-time dynamics, while its accuracy brakes down  at sufficiently short-times.
This type of failure  represents a common feature to all statistical or quantum effective field theories. For example, the multipole expansion for the electric field generated by a finite charge distribution becomes highly inaccurate  at short distances, of the order of the charge distribution size. 

\section{The Effective Langevin Equation} 
\label{ESTmd}

The EST presented in the previous section provides a low time resolution path integral representation of the stochastic dynamics generated by the LE  (\ref{lang}), in which the fast dynamics has been integrated out and replaced by a new effective interaction term, $V_{eff}^{R}({\bf X})$ involving only the slow components of the paths.  
The  goal of this section is to prove that such a low time resolution dynamics can be equivalently formulated in terms of an effective Langevin Eq., in which the effects of the fast modes is implicitly taken into account through a systematically calculable correction to the diffusion coefficient. To this goal, we first formulate an ansatz for such an 
effective Langevin Eq., and then determine a condition which follows from requiring that our equation should generate the path integral (\ref{EST}),  which defines the EST.
  
   Let us therefore discuss the stochastic dynamics generated by the Eq.
\be
\label{theory}
\dot{\bf X} = -\frac{D({\bf X}_i)}{k_B T}~\nabla U({\bf X}_i) + (1-\alpha) \nabla D({\bf X}_i) + \sqrt{2 D({\bf X})} ~{\bf\eta}(t),
\ee
where  
 $\alpha$ is a insofar undefined constant whose origin will be discussed shortly,  $\eta(t)$ is the usual delta-correlated Gaussian noise and $D({\bf X})$ is a position-dependent diffusion coefficient in the form
\be
D({\bf X}) = D_0 + d({\bf X}).
\ee 
In the following, the function $d({\bf X})$ and its  Laplacian will be assumed to be small, compared to the corresponding hard scales:
\be
\label{expansion}
\frac{d({\bf X})}{D_0}\ll 1, \qquad \frac{\nabla^2 d({\bf X})}{ b\Omega} \ll 1. 
\ee 

We shall now determine a condition on the function $d({\bf X})$ which follows from requiring that arbitrary time-dependent averages of configuration-dependent  observables generated by Eq. (\ref{theory}) coincide with those calculated in the EST, i.e. using the renormalized path integral Eq.~(\ref{EST}). 
Note that, in the right hand side of Eq.  (\ref{theory}) we have introduced the term $(1-\alpha) \nabla D({\bf X})$, which does not appear in the ordinary LE~(\ref{lang}).  In order to clarify the motivation for such a term, we need to make a short digression on the so called "Ito-Stratonovich dilemma", which affects stochastic differential Eq.s with multiplicative noise. Following closely the discussion in \cite{colored}, let us consider a generic stochastic differential Eq. in the form
 \be
\label{multiplicative}
 \dot {\bf X} = {\bf f}({\bf X}) + g({\bf X})~ \eta(t).
 \ee 
In such a family of Eq.s,  an ambiguity arises from the fact that the value of the integral which appears in its formal solution, 
\be
\label{integral}
{\bf \mathcal{J}}(t, \Delta t) \equiv \int_t^{t+\Delta t} ds ~ g[{\bf X}(s)] ~{\bf \eta}(s),
\ee
depends on wether the function $g({\bf X}(s))$ is evaluated before or after the action of the random force~${\bf \eta}(t)$. 
The most general definition of the integral (\ref{integral}) can be cast in the form
\be
\label{integral2}
{\bf \mathcal{J}}_\alpha(t, \Delta t) \equiv g[\alpha {\bf X}(t+\Delta t)+ (1-\alpha){\bf X}(t)]~\int_t^{t+\Delta t}~ds~ {\bf \eta}(s),
\ee
where the  real number $0\le \alpha\le 1$ specifies the prescription used to define the stochastic Calculus. For example, $\alpha=0$ leads to an Ito Calculus, while $\alpha=\frac{1}{2}$ corresponds to a Stratonovich Calculus.

For a generic Eq. in the form (\ref{multiplicative}),  the ambiguity (\ref{integral2}) is not resolved in the continuum limit. That is to say that different choices of $\alpha$ lead in general to different Fokker-Planck Eq.s.
However, in the specific case of the Eq. (\ref{theory}), the ambiguity is resolved by the addition of the term $(1-\alpha)~\nabla D({\bf x})$, which  assures that, for any choice of prescription $\alpha$, the resulting probability density obeys the same Fokker-Planck Eq.~\cite{colored}, 
\be
\frac{\partial}{\partial t} P({\bf X}, t) = \nabla \left[D({\bf X}) ~\left(\frac{1}{k_B T}~\nabla U({\bf X}) + \nabla~\right)~ P({\bf X}, t)\right].
\ee
In addition, it can be shown that the solution of such a Fokker-Planck Eq. converges to the correct 
Boltzmann's distribution, in the long-time limit\cite{colored}:
\be
 P({\bf X},t) \stackrel{t\to\infty}{\rightarrow} ~ \textrm{const.}~\times~ \exp\left[-\frac{U({\bf X})}{k_B T}\right]. 
\ee
 
While the Fokker-Planck Eq. associated to Eq. (\ref{theory}) is independent on the choice of $\alpha$, the path integral representation of the conditional probability  $P({\bf X}_f, t| {\bf X}_i,0)$ depends  such a parameter
 and reads
\be
P({\bf X}_f, t| {\bf X}_i,0) &=& \int \hat{\mathcal{D}}{\bf X}~ \exp \left[-\int_0^t~d\tau~\frac{1}{4 D({\bf X})} \left(\dot {\bf X} + \frac{D({\bf X})}{k_B T}~\nabla U({\bf X}) - 
(1-2\alpha)~ \nabla D({\bf X}) ~\right)^2  \right.\nonumber\\
&&\left. + \alpha \nabla \cdot  \left(-\frac{\Delta t~D({\bf X})}{k_B T}~\nabla U({\bf X}) +  \nabla D({\bf X}) \right)~\right],
\ee
where $\hat{\mathcal{D}} {\bf X}$ is a position-dependent measure:
\be
\label{measure}
\hat{\mathcal{D}} {\bf X} \equiv \prod_{i=1}^{N_t}~d{\bf X}_i~\left(\frac{1}{4 \pi \Delta t D({\bf X}_i)}\right)^{\frac{3 N_p}{2}}.
\ee

In the following, we shall adopt the Ito convention $\alpha=0$, in  which the stochastic differential Eq.  (\ref{theory}) is defined by the rule:
\be
\label{main}
{\bf X}_{i+1} = {\bf X}_i -\frac{\Delta t~D({\bf X}_i)}{k_B T}~\nabla U({\bf X}_i) + \Delta t~ \nabla D({\bf X}_i) + \sqrt{2 D({\bf X})~\Delta t} ~{\bf X}_i.
\ee
The periodic path integral generated by such an Eq. is
\be
\label{PIR}
Z_{ELE}(t) &\simeq&\prod_{i=1}^{N_t}~\int d({\bf X}_i~\left(\frac{1}{4 \pi \Delta t D({\bf X}_i)}\right)^{\frac{d N_p}{2}}.~\exp \left[-\sum_i ~\frac{({\bf X}_{i+1}-{\bf X}_{i})^2}{4 D({\bf X}_i) \Delta t}~
+ \frac{({\bf X}_{i+1}-{\bf X}_{i})}{ 2 k_B T}  \nabla U({\bf X}_i)   + \frac{D({\bf X}_i)}{4 (k_B T)^2} (\nabla U({\bf X}_i))^2\Delta t \right.\nonumber\\
&&\left.  -\frac{({\bf X}_{i+1}-{\bf X}_{i})}{2 D({\bf X}_i)} \nabla D({\bf X}_i)+   \frac{1}{4~D({\bf X}_i)} (\nabla D({\bf X}_i))^2\Delta t - \frac{ \Delta t}{2 k_B T} \nabla U({\bf X}_i) \cdot \nabla D({\bf X}_i)\right].
\ee

We now show that  the terms appearing in the second line provide sub-leading contributions in the expansion scheme (\ref{expansion}). To this end, we  first observe that  the Langevin Eq. (\ref{theory}) implies that, on  average,     
\be
-\frac{\Delta t}{2 k_B T}\nabla U({\bf X}_i) \cdot \nabla D({\bf X}_i) \simeq \frac{1}{2 D({\bf X}_i)} ({\bf X}_{i+1}- {\bf X}_i),
 \cdot \nabla D({\bf X_i}) -\frac{\Delta t}{2 D({\bf X}_i)} \nabla D({\bf X}_i)^2.
\ee 
Substituting this into Eq. (\ref{PIR}) we find
\be
\label{PIR4}
Z_{ELE}(t) &\simeq&\prod_{i=1}^{N_t}~\int d({\bf X}_i~\left(\frac{1}{4 \pi \Delta t D({\bf X}_i)}\right)^{\frac{d N_p}{2}}.
~\exp \left[-\sum_i ~\frac{({\bf X}_{i+1}-{\bf X}_{i})^2}{4 D({\bf X}_i) \Delta t}~
+ \frac{({\bf X}_{i+1}-{\bf X}_{i})}{ 2 k_B T}  \nabla U({\bf X}_i)   \right.\nonumber\\
&&\left. + \frac{D({\bf X}_i)}{4 (k_B T)^2} (\nabla U({\bf X}_i))^2\Delta t -   \frac{1}{4~D({\bf X}_i)} (\nabla D({\bf X}_i))^2\Delta t \right].\nonumber\\
\ee
Taking the continuum limit, and recalling the fundamental theorem of the Ito Calculus --- Eq.~(\ref{ito})--- we obtain
\be
Z_{ELE}(t) &=&  \oint \hat{\mathcal{D}}{\bf X} ~\exp \left[-\int_0^t~d\tau~
~\frac{\dot {\bf X} ^2}{4 D({\bf X})}
+ \frac{D({\bf X})}{4 (k_B T)^2}~(\nabla U({\bf X}))^2 - \frac{D({\bf X})}{2 k_B T}\nabla^2 U({\bf X}) 
 - \frac{1}{4~D({\bf X})} (\nabla D({\bf X})^2)   \right]\\
&=& \oint \hat{\mathcal{D}}{\bf X} ~\exp \left[-\int_0^t~d\tau~
~\frac{\dot {\bf X} ^2}{4 D({\bf X})}
+ \frac{D({\bf X})}{4 (k_B T)^2}~(\nabla U({\bf X}))^2 - \frac{D({\bf X})}{2 k_B T}\nabla^2 U({\bf X}) -
\frac{1}{4}\nabla^2~D({\bf X}) + \frac{D({\bf X}_i)}{4} \nabla^2 \log\frac{D({\bf X})}{D_0} \right],\nonumber\\
\ee

which, to leading order in $d({\bf X})$, reads 
\be
Z_{ELE}(t) &=&  \oint \hat{\mathcal{D}}{\bf X} ~\exp \left[-\int_0^t~d\tau~
~\frac{\dot {\bf X} ^2}{4 D({\bf X})}
+ \frac{D({\bf X})}{4 (k_B T)^2}~(\nabla U({\bf X}))^2 - \frac{D({\bf X})}{2 k_B T}\nabla^2 U({\bf X}) +
 \frac{d({\bf X})}{4 D_0} \nabla^2 d({\bf X})~\right].
\ee
The last term in the exponent is the contribution coming from the second line of Eq. (\ref{PIR})  and is  of  higher order in the expansion scheme (\ref{expansion}) and therefore can be neglected.

Finally, using Einstein's relationship, it is immediate to show that  to the same order in our  expansion scheme, the correction to the freely diffusive term $\frac{\dot {X}^2}{4 (D_0 + d({\bf X}))}$ cancels out with  the corresponding correction to the  measure 
\be
\hat{\mathcal{D}} {\bf X} \equiv \prod_{i=1}^{N_t}~d{\bf X}_i~\left(\frac{1}{4 \pi \Delta t(D_0+ d({\bf X}_i))}\right)^{\frac{3 N_p}{2}}.
\ee
Hence, we arrive to the compact expression for the periodic path integral associated to the effective Langevin Eq. defined by Eq. (\ref{main}). 
\be
\label{final}
Z_{ELE}(t) &\simeq&  \oint  \mathcal{D}{\bf X} ~\exp \left[-\int_0^t~d\tau~
~\frac{\dot {\bf X} ^2}{4 D_0}
+ V_{eff}({\bf X}) + d({\bf X}) V_{eff}({\bf X}) \right].
\ee

We now impose the condition that the effective Langevin Eq.  (\ref{main}) should generate the same  stochastic dynamics of the EST, i.e. the same periodic path integral:
\be
Z_{ELE}(t) \equiv Z_{EST}(t), \qquad \forall t.
\ee
This condition can be re-written as
\be
\label{equi}
\langle \exp\left[-\int_0^t~d\tau~ V_{eff}({\bf X}) ~d({\bf X}) \right] \rangle \equiv \langle ~\exp\left[-D_0~ \int_0^t~d\tau ~V_{eff}^R({\bf X})\right]\rangle,
\ee 
where the notation $\langle \cdot \rangle$ denotes the average performed over the ensemble of periodic paths generated by the ordinary LE (\ref{lang}).  
Since such an Eq. must hold for any  total time interval $t$, then 
\be
\label{equi2}
\langle \exp\left[-V_{eff}[{\bf X}(\tau)] ~d[{\bf X}(\tau)] \right] \rangle \equiv \langle \exp\left[-D_0~V_{eff}^R[{\bf X}(\tau)]\right]\rangle, \qquad \forall \tau\in[0,t].
\ee 

Now we recall that the renormalized part of the potential $V_{eff}^R({\bf X})$ and the renormalized part of the diffusion coefficient $d({\bf X})$ provide small corrections  to the  stochastic dynamics generated by the ordinary LE. Hence, we can expand the exponentials to leading order and obtain:
\be 
\label{equi3}
\langle  V_{eff}({\bf X}(\tau)) ~d({\bf X}(\tau))  \rangle \equiv  D_0~\langle V_{eff}^R({\bf X}(\tau))\rangle, ~ \qquad \forall \tau\in[0, t].
\ee

Eq. (\ref{equi3}) represents a  condition on the correction term to the diffusion constant $d({\bf X})$ which is sufficient to ensure that the stochastic differential Eq. (\ref{main})  generates  the low time resolution dynamics  of the EST.  
The main assumptions made to derive it are the existence of a separation of time scales, and the related expansion scheme (\ref{expansion}). 

Now we make one further approximation, which consists in  describing  the correction to the diffusion coefficient $d({\bf X})$ in the mean-field approximation, i.e. 
\be 
\label{equi4}
\langle  V_{eff}({\bf X}(\tau)) ~d({\bf X}(\tau))  \rangle \simeq \langle  V_{eff}({\bf X}(\tau))\rangle ~\langle d({\bf X}(\tau))  \rangle,
\ee
which implies  
\be
\label{MFd}
\langle~ d(\tau))~\rangle  \simeq D_0~\frac{ \langle~ V_{eff}^R({\bf X}(\tau))~ \rangle}{ \langle~ V_{eff}({\bf X}(\tau))~ \rangle}.
\ee 
We stress the fact that the averages involved in Eq.(\ref{MFd}) is performed over all periodic paths generated by the ordinary LE. It depends on time,  but does not depend on the position. We also note that Eq.(\ref{MFd})  diverges if the \emph{average value} of $V_{eff}({\bf X}(\tau))$ vanishes at some time $\tau$. In principle, this problem may be cured by adopting some regularization prescription. However, in practice, for all systems we have considered,  the average of the effective potential was always found to be a negative number, for all times $\tau$. This is because the stochastic trajectories are most likely to visit regions of configuration space where the force is small, and the effective potential is dominated by the Laplacian contribution ---cfr. Eq. (\ref{Veff})---.

Within the mean-field approximation for the diffusion coefficient, the Ito Eq.  (\ref{main}) which defines the effective Langevin Eq.  reduces to one with a non-multiplicative noise:
\be
\label{ELEMF}
{\bf X}_{i+1} = {\bf X}_{i} - \frac{\Delta t ~(D_0 + \langle d(i)\rangle)}{k_B T}~\nabla U({\bf X}_i) +  \sqrt{2 (D_0 + \langle d(i)\rangle)~\Delta t}~{\bf\eta}_i.
\ee

From Eq. (\ref{ELEMF}) it is manifest that, in the mean-field approximation (\ref{MFd}) and to the lowest order in the expansion scheme (\ref{expansion}), the dynamics of the fast modes can be integrated out  by means of a time-dependent rescaling of the time intervals:
\be
\label{rescale}
D_0 (t_{i+1}-t_i)~\rightarrow ~(D_0 +\langle d(i)\rangle)~(t_{i+1}-t_i),
\quad\Rightarrow \quad(t_{i+1}-t_i) = \frac{1}{1 +\langle d(i)\rangle}~\Delta t.
\ee   

Based on such observation, we are finally in a condition to define a simple three-step algorithm,  which yields the stochastic dynamics at low time resolution power:
\begin{enumerate}
\item One generates an ensemble of stochastic paths by integrating the LE, using a  \emph{large} time step $\Delta t_L= \Delta t/b$. Here, $\Delta t$ is a small integration time step interval for which the LE is convergent. This means that the dynamics which occurs at time scales smaller than $\Delta t$ is not expected to contribute to the process under investigation.  
\item Such paths are used to compute the average time-dependent correction to the diffusion coefficient $\langle d(\tau)\rangle$, according to Eq. (\ref{MFd}).
The frequency cut-off in the effective theory, $b\Omega$, which enters in the definition of the renormalized effective potential $V_{eff}^R({\bf X})$ --- cfr.  Eq. (\ref{L-leq-3})--- is related to the \emph{large} integration time step by
\be
b \Omega = b~\frac{2 \pi}{ \Delta t} = \frac{2 \pi}{ \Delta t_L}. 
\ee
\item Each time intervals between each consecutive instants $t_{i+1}$ and $t_i$ are rescaled according to Eq. (\ref{rescale}), i.e.  by the dilation factor provided by the renormalized average diffusion coefficient. 
\end{enumerate}
It is important to note that, in general,  the \emph{average value} of the configuration-dependent operators is expected to vary over time scales which are of the order of the thermal equilibrium relaxation time, i.e. typically several orders of magnitude larger than the time scales associated to the slowest local microscopic conformational changes.  
This means that, in numerical simulations,  the value of the mean-field  correction to the diffusion coefficient, $\langle d(\tau)\rangle$ evolves very slowly, hence it needs to be updated only after a large number of elementary integration time steps. 
In the following, we shall refer to this algorithm as to the effective Langevin Eq. (ELE) approach. 

 \begin{figure}[t!]
		\includegraphics[clip=,width=8 cm]{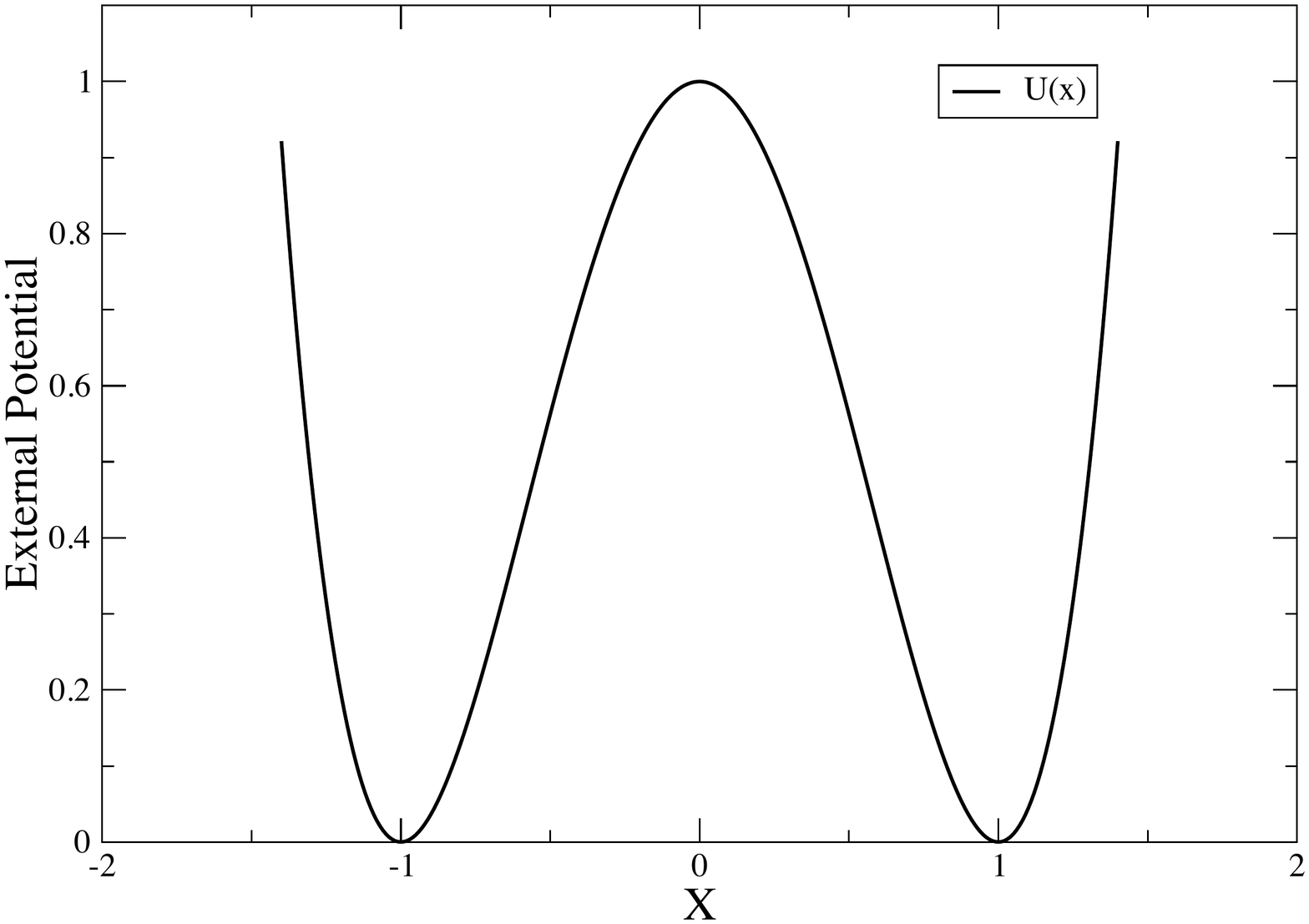}
	\includegraphics[clip=,width=8 cm]{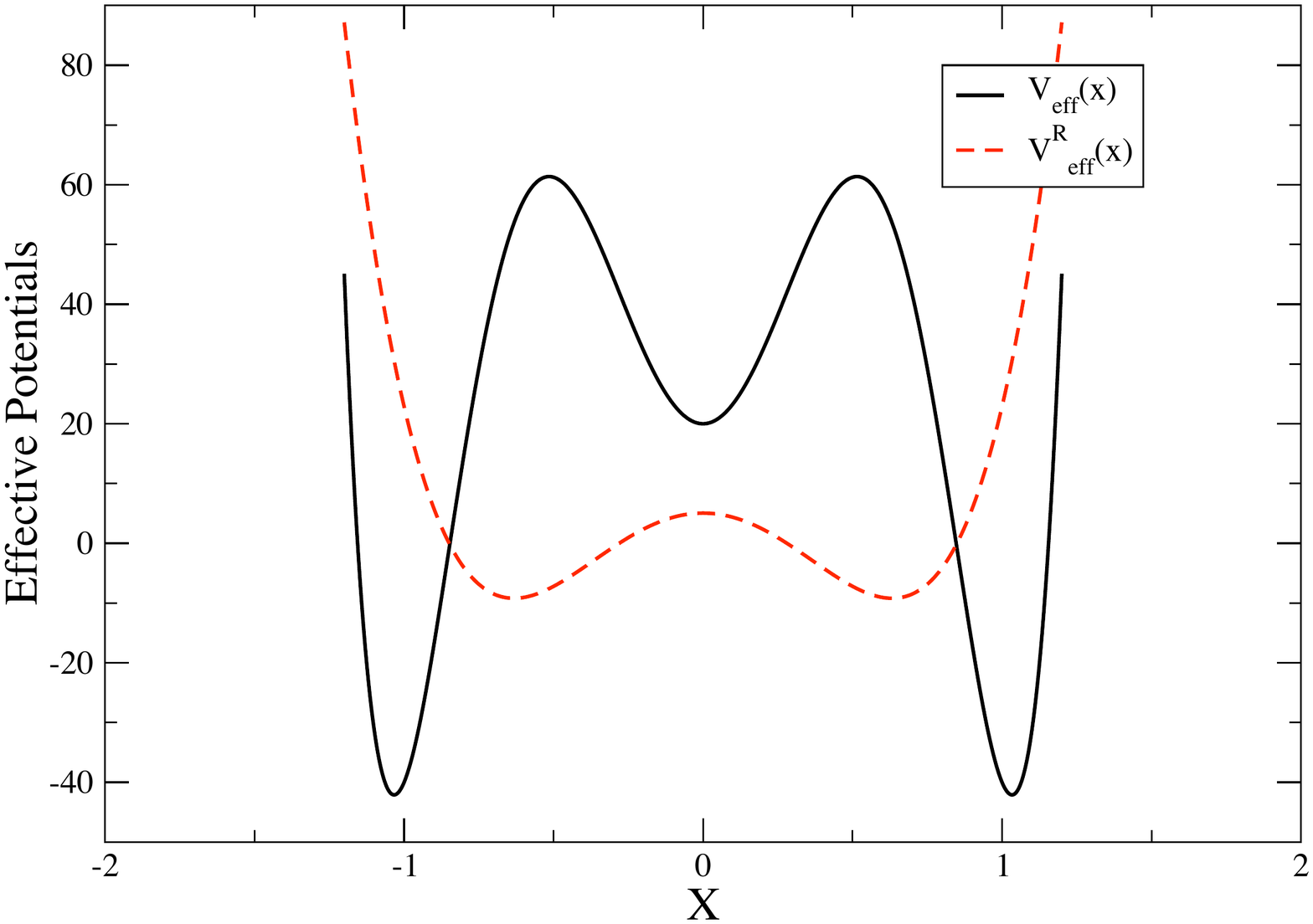}		
	\caption{Left panel: the double well-potential of the one-dimensional toy model. Right panel: the corresponding L=0 (i.e. bare) and  L=1 contributions to the effective potential (The L=1 contribution is not on scale)}
	\label{ELEvsOLE1Dpots}
\end{figure}

\section{Two Illustrative Applications}
\label{illustrative}

In this section, we illustrate and test the ELE approach proposed in the previous section. We begin by discussing  a simple toy model, consisting of a point-particle diffusing in an external one-dimensional potential, and then apply the same approach to investigate the unfolding  of a small protein, at a high temperature.

\subsection{Diffusion in a one-dimensional double-well}
\begin{figure}[t!]
	\includegraphics[clip=,width=14 cm]{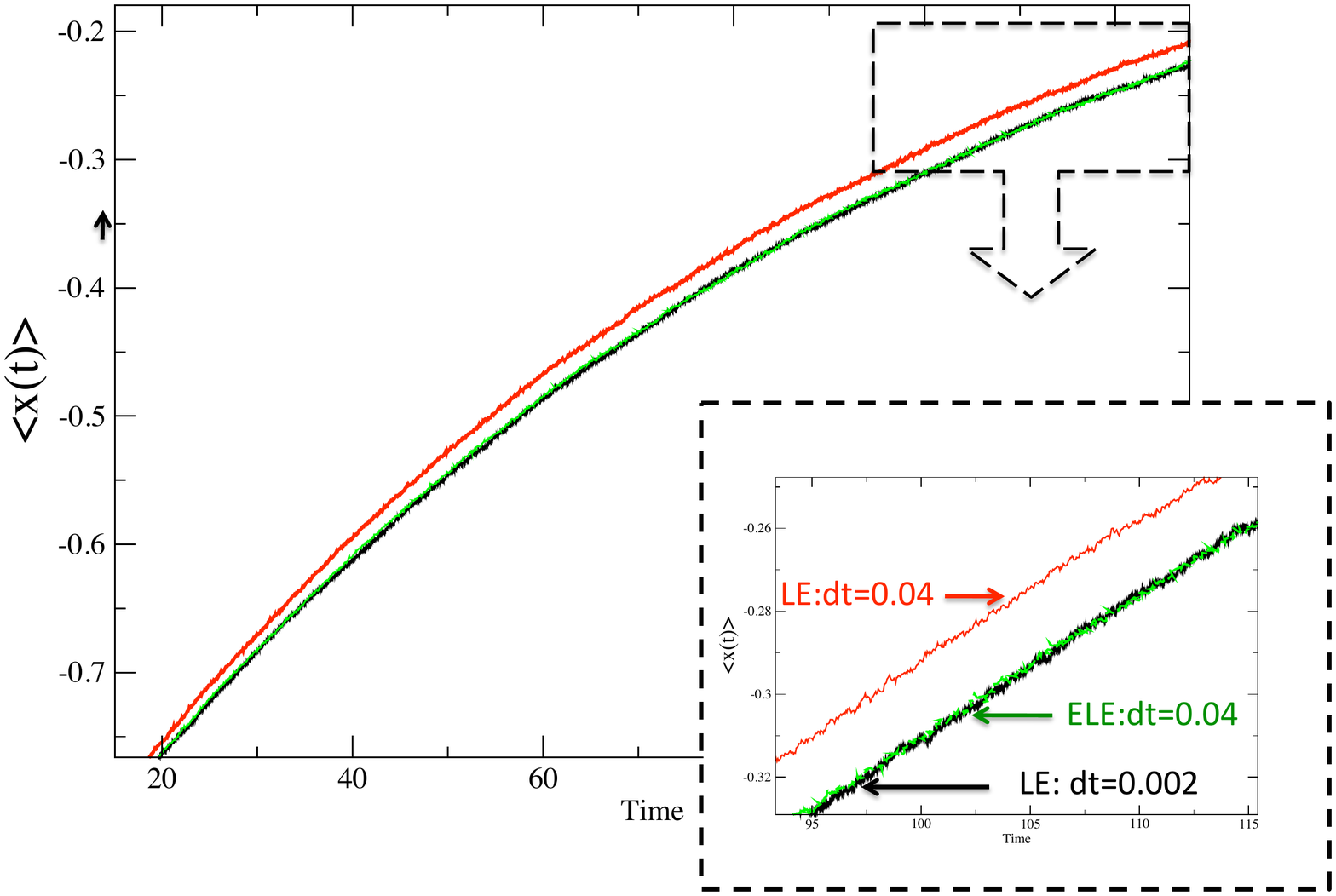}
		\includegraphics[clip=,width=14 cm]{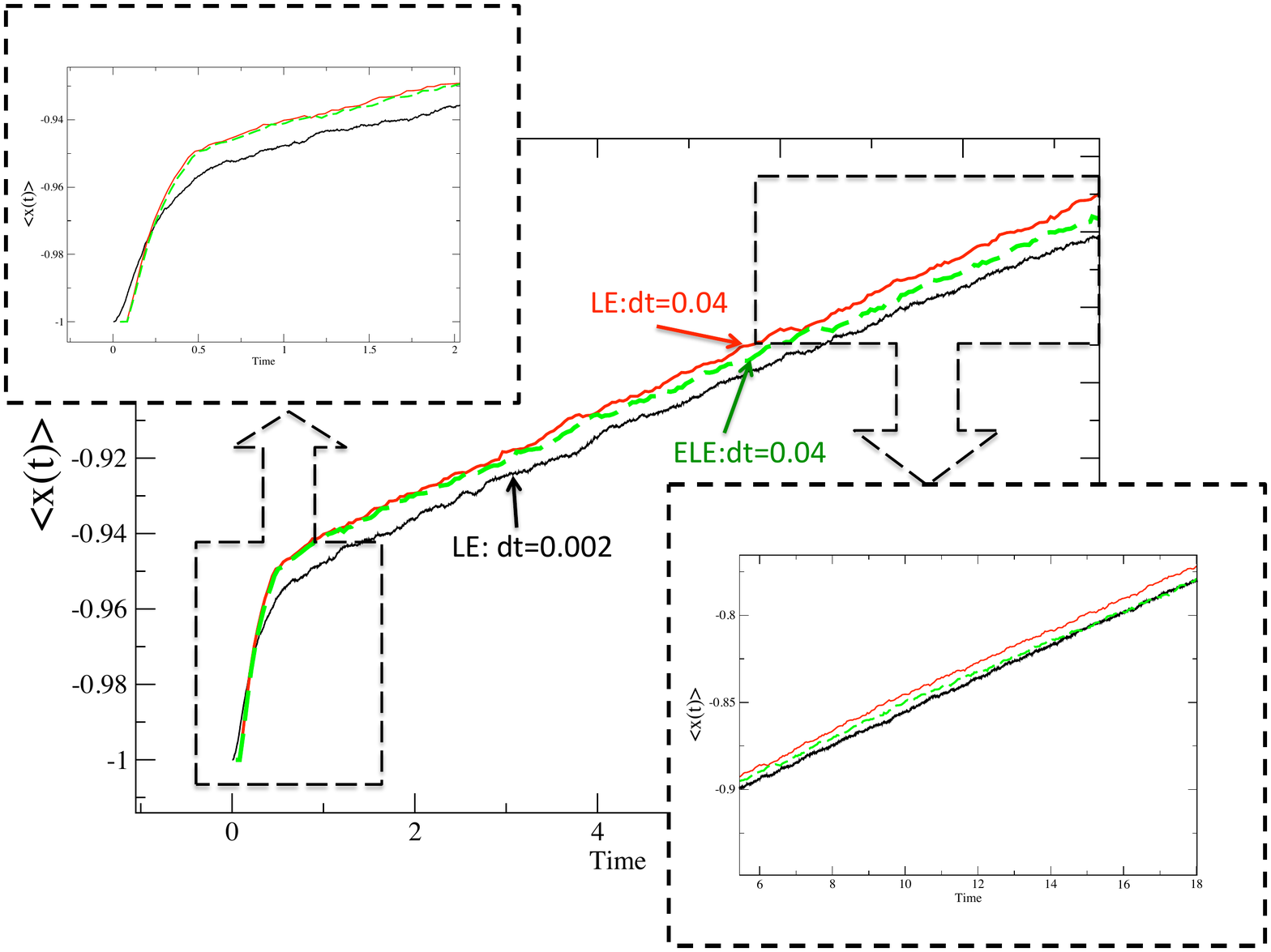}
			\caption{Time evolution of the average position of a point particle  diffusing in a one dimensional double-well potential. The upper panel represents the time evolution  at long times, when the ELE is expected to be accurate. The lower panel displays  the evolution at short times, where the ELE becomes inaccurate. The red line is the result of the integration of the ordinary LE with a large integration step $\Delta t_L=0.04$, the black line is the result of integrating the LE with a small  time step $\Delta t=0.002$ and the green line is the result of solving  ELE  with the large integration time step $\Delta t_L=0.04$. }
			\label{ELEvsOLE1Dres}	
\end{figure}

Let us consider the diffusion of a point-particle in the external double-well potential
\be
\label{U1D}
U(x) = \alpha (1-x^2)^2.
\ee
We have chosen a system of units in which $\alpha=1$, $D_0=1$ and $\beta=1/(k_B T) = 5$. The potential $U(x)$,  the corresponding effective potential $V_{eff}(x)$ and the  renormalized effective potential $V_{eff}^R(x)$ evaluated to order $L=1$ for this system are plotted in Fig.\ref{ELEvsOLE1Dpots}.
The dynamics of such a system is characterized by a decoupling of time scales, since the quasi-free diffusion in the bottom of the wells is much slower than the crossing of the transition regions, where the force is large.  

Our goal is to compare the predictions for the time evolution of the average position $\langle x(t)\rangle$ obtained in the effective theory defined by the ELE and in the original theory, defined by the ordinary LE.  We have generated two sets of  90,000 trajectories, with initial condition, $x(0)=-1$. The two  ensembles of trajectories were obtained by integrating the  LE (\ref{lang}),  using two different elementary time steps: a "small" one, $\Delta t=0.002$, and a "large" one, $\Delta t_L=0.04$.  Notice that $\Delta t $ is  20 times smaller than $\Delta t_L$.  
The results are reported in  Fig.~\ref{ELEvsOLE1Dres}, where the time evolution of $\langle x(t)\rangle$ is shown at short times (lower panel) and long times (upper panel). Both panels show that the results of the straightforward integration of the LE using the large integration time step $\Delta t$ are inconsistent with those obtained integrating the same equation, using the small integration time step $\Delta t_L$. Hence,  the short-time dynamics which is  cut-off by the large time step  cannot be neglected. 

On the other hand, in the ELE, such a fast dynamics is not neglected, but it is effectively taken into account at the mean-field level, through the time-dependent  correction to the diffusion constant $\langle d(t)\rangle$. Such a term was calculated from  Eq.(\ref{MFd}) using the Langevin trajectories obtained  with the large time step, and  is plotted in Fig.~\ref{Vpots}. We see that the integration of the fast modes leads to an effective slow-down of the dynamics. Indeed, in the ELE all the time steps  become approximatively 
 $5\%$ longer.  In the upper panel of Fig. ~\ref{ELEvsOLE1Dres} we can see that, after applying the rescaling transformation (\ref{rescale}),   one recovers an excellent  agreement with the predictions obtained integrating directly the LE with the small time step. 
 \begin{figure}[b]
	\includegraphics[clip=,width=8 cm]{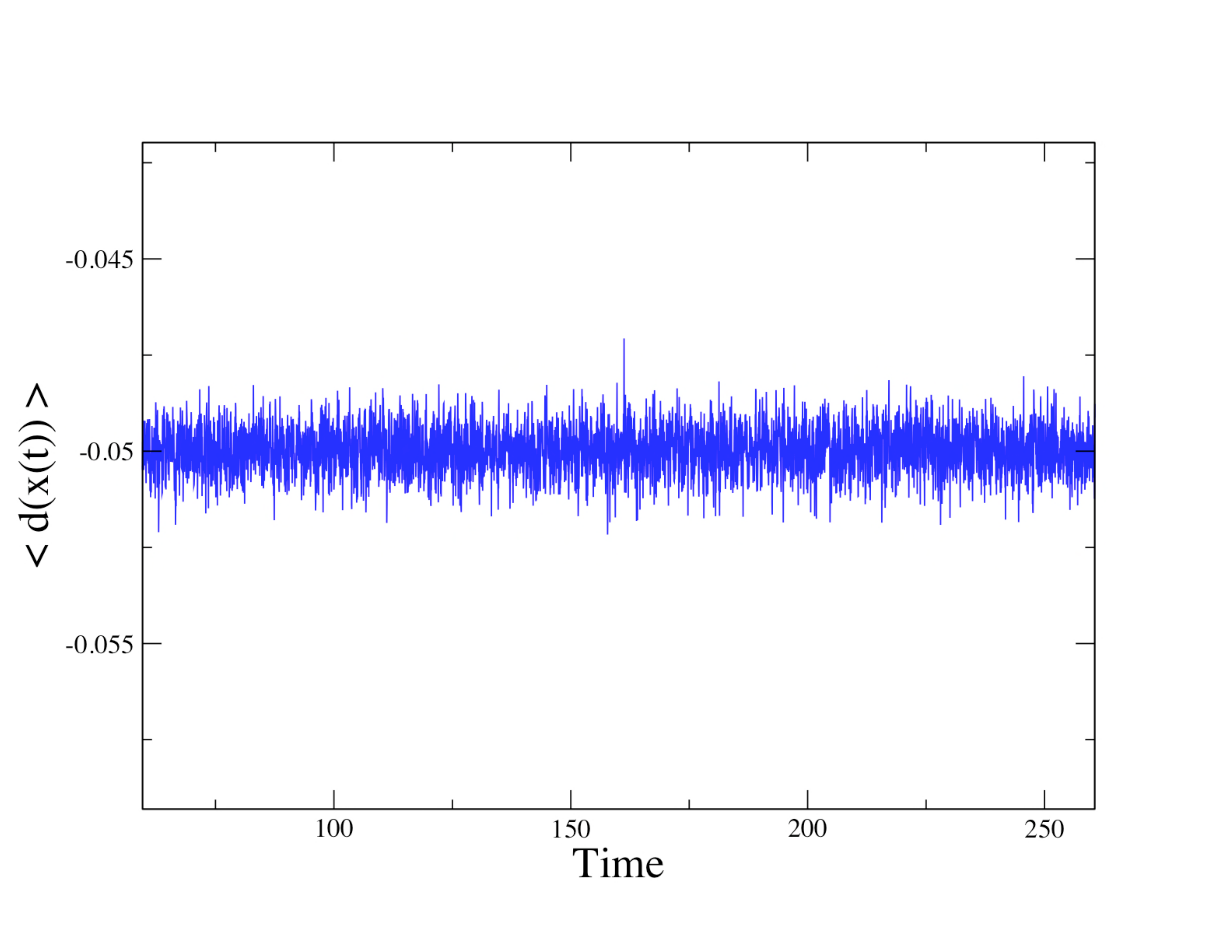}		
	\caption{Calculated time evolution of the  $\langle d(x(t)) \rangle$ in the one-dimensional toy model define by Eq. (\ref{U1D}). }
\label{Vpots}
\end{figure}

We emphasize again that all effective field theories (including our EST, or equivalently the ELE) are expected to accurately describe only the long-time (i.e.  infra-red) dynamics. They are  not expected to provide reliable descriptions of the time evolution of the system in the short-time (i.e. ultra-violet) regime. This feature is clearly evident in the lower panel of   ~\ref{ELEvsOLE1Dres}, where we show the evolution of $\langle x(t)\rangle$ at short times. We see that for $t\lesssim 10$ the results of the ELE calculation obtained with large integration time step $\Delta t_L$ start to deviate from the results obtained from the LE, with the small integration time step $\Delta t$. This is regime where our effective theory breaks down. On the other hand, for $t\gtrsim 15$ the ELE curve approaches the LE results obtained with the small integration time step. In such a regime, the ELE provides an excellent description of the dynamics.

\subsection{High temperature protein denaturation.}

As a second test of the ELE approach, we study the  unfolding of the 16-residue  $C$-terminus of protein GB1 shown in Fig.\ref{GB1}. This system  was used as a test system in our previous studies, in the dominant reaction pathways approach \cite{DRP2}.
We adopt a coarse-grained Go-type model \cite{go}, in which the explicit degrees of freedom are beads which represent the single  amino-acids. 
The energy function of this model is assumed to be the sum of pair-wise interactions: 

 \begin{figure}[t!]
	\includegraphics[clip=,width=10 cm]{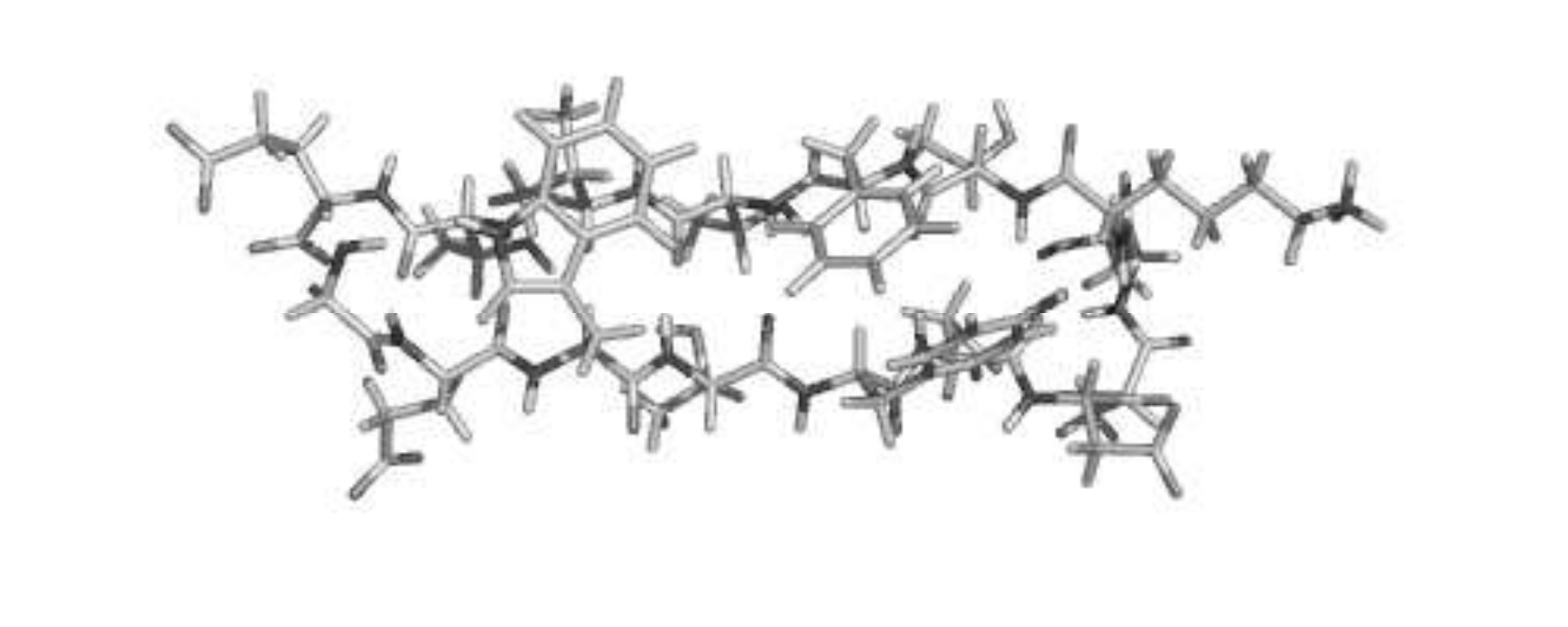}		
	\caption{16-residue C-terminus of protein G-B1 (PDB code 2gb1).}
	\label{GB1}
\end{figure}

\be
\hspace{-2 cm}U({\bf X}) = \frac{1}{2} \sum_k  k (|{\bf r}_{k+1}-{\bf r}_k|-a)^2
+ \frac 12\sum_{i\ne j} 4 \epsilon \left[\left(\frac{\sigma}{|{\bf r}_{j}-{\bf r}_i|}\right)^{12} - G_{ij}~\left(\frac{\sigma}{|{\bf r}_{j}-{\bf r}_i|}\right)^{6}\right]
\ee
$a=0.38$~nm represents the average distance between two consecutive $\alpha$-carbons and $k=1000 ~\textrm{kJ mo}l^{-1} \textrm{nm}^{-2}$ is the elastic constant of the harmonic spring.  The strength of the Lennard-Jones attraction is set by the parameter $\epsilon= 4 ~\textrm{kJ mol}^{-1}$, while $\sigma=0.3~$nm represents an effective residue size.  $G_{ij}$ is the matrix of native contacts, i.e. $G_{ij}$  is set to $1$ if the distance between the residues $i$ and $j$ in the native conformation is less than $0.65$~nm, and  $0$ otherwise. 

In the left panel of Fig. \ref{Mol1} we show the average time evolution for $200$~ps, of the fraction of native contacts of this chain, starting from a configurations close to the experimentally measured native configuration.  
The average was performed over 900 independent trajectories, generated by integrating the LE with a time step $dt=0.002$~ps, at a temperature $T=200 K$ and diffusion constant $D_0=0.8 \textrm{nm}^2 \textrm{ps}^{-1}$.   
We see that, at such a low temperature, the experimentally measured native state remains stable.  On the other hand, at high temperatures, the native structure is thermodynamically unstable and the protein spontaneously unfolds. 

We computed the time evolution of the average fraction of native contacts, during the high temperature unfolding reaction. The purpose of this section is to  assess the validity of the ELE method, by comparing the results of the microscopic  calculation obtained from the LE with a small integration time step $\Delta t= 0.002$~ps
with those obtained from the  ELE, using a 20-fold larger  integration time step, $\Delta t_L=0.05$~ps. 
 We have observed that, for time steps smaller than $0.002$~ps, the results of the Langevin simulations are independent on the choice of the integration time step, within the statistical errors. 

In  analogy with the  previous one-dimensional example, we have generated two sets of 900 independent trajectories at $T=300 K$ and $D_0=1.2 \textrm{nm}^2 \textrm{ps}^{-1}$, by integrating the LE with the small and large time steps,  starting from the experimentally determined native state. Fig. \ref{Mol1} shows that a small yet statistically significant discrepancy is observed between the results of the LE with time steps $\Delta t$, and $\Delta t_L$. 
Indeed, the  results of the LE obtained with the large integration time step fall consistently short of the corresponding  points obtained with the small integration time step.  As in the previous one-dimensional example, this is a clean signature of the fact  that the dynamics which occurs at the time scale of $10^{-2}$~ps cannot be simply cut-off. 
However we can see that, once such a dynamics is effectively taken into account through  the ELE,  the agreement with the long-time LE  results obtained using the small integration time step is recovered. We also note that our effective theory breaks down in the short-time regime, as expected.  However, such a limitation of the ELE method does not represent a problem, in practical applications, since the short-time molecular dynamics can be very efficiently simulated using the existing algorithms.  

 \begin{figure}[t!]
	\includegraphics[clip=,width=8 cm]{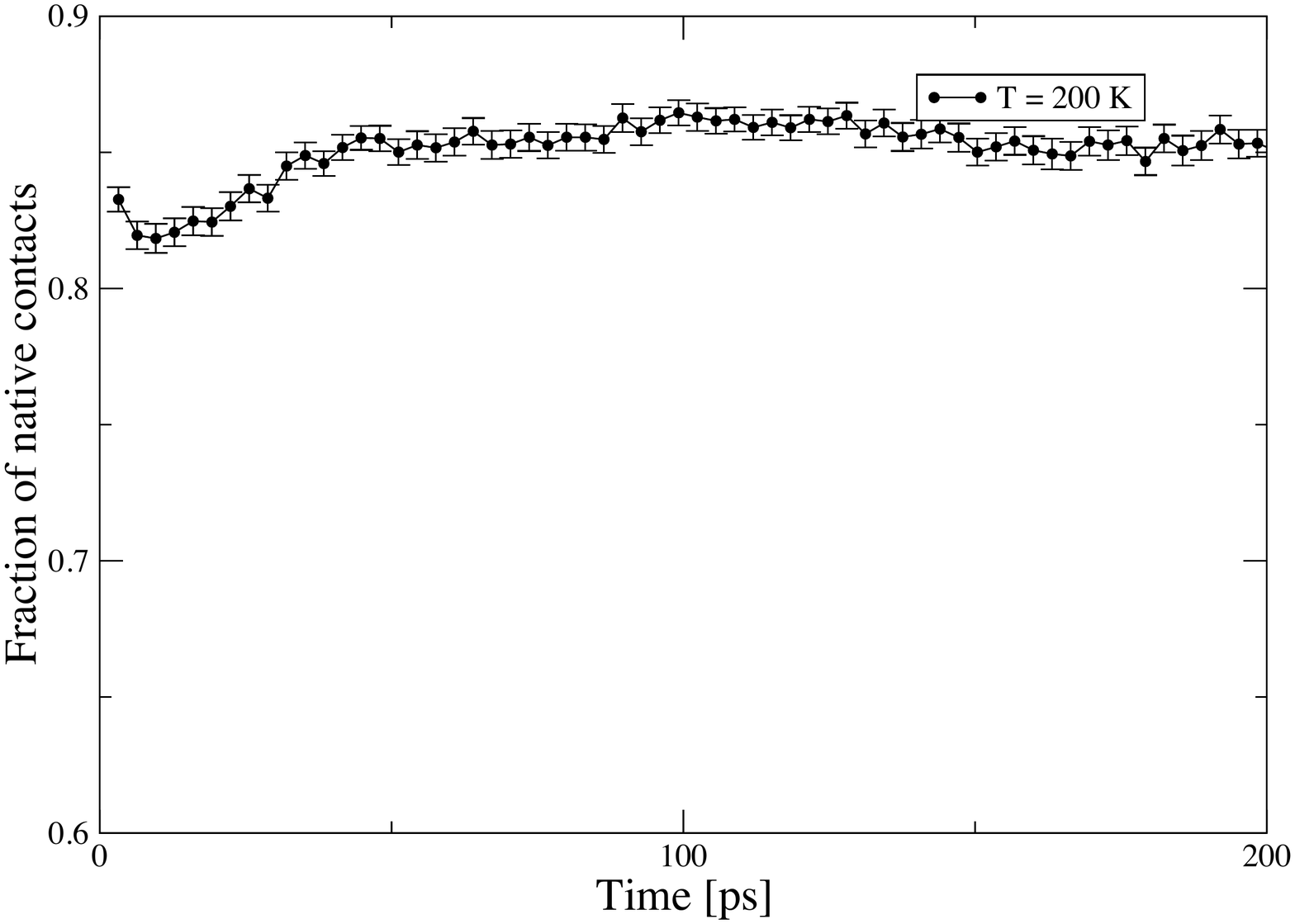}
	\includegraphics[width=8cm]{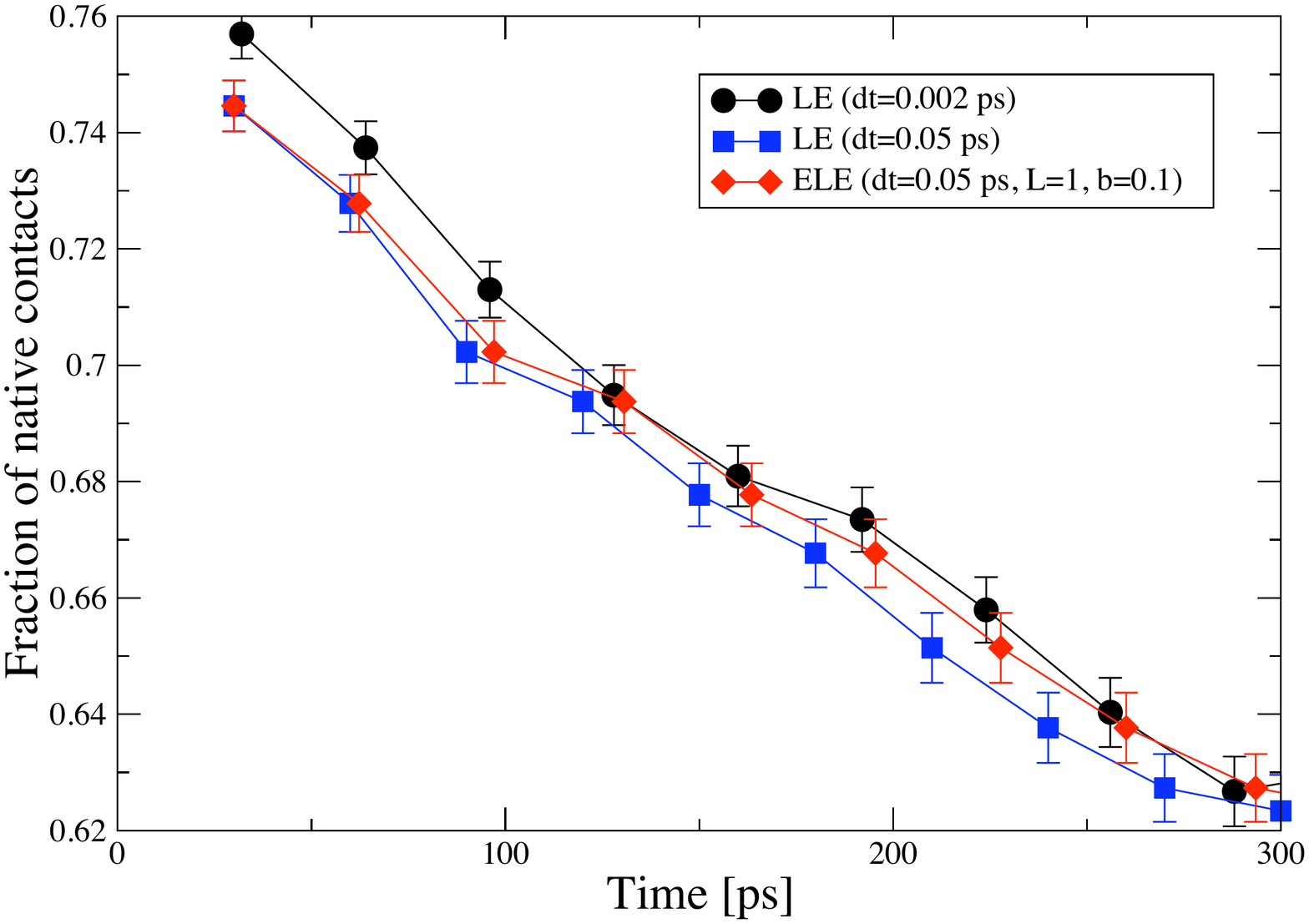}\\
	\caption{Left panel: Time evolution of the fraction of native contacts of the $\beta$-hairpin, at the temperature $T=200 K$. Right panel:Comparison of the results of the LE  and ELE simulations for the time evolution of the fraction of native contacts, for the 16-residue poly-peptide chain shown in Fig. \ref{GB1}.}
	\label{Mol1}
\end{figure}

\section{Conclusions}
\label{conclusions}

In this work we have introduced an effective theory based on a first-order stochastic differential equation which describes the microscopic molecular dynamics, at a
low time resolution power.  In such an approach, the effects of the fast dynamics which is excluded by using a large integration time step are implicitly accounted for by means of an effective time-dependent diffusion constant, which is derived using the RG approach.  
The assumptions used in the derivation are: (i) the existence of a gap in the internal dynamical time scales of the system and (ii) the mean-field approximation in the calculation of the effective diffusion constant.

We have illustrated and validated our method by studying the diffusive dynamics  of a one-dimensional toy-model and of a simple coarse-grained model for a protein fragment. 
In both cases, we have found that our effective theory yields the correct long-time dynamics, even when one uses an integration time step which is 20 time larger than the one used in the ordinary Langevin  simulations. 
  
The present preliminary study did not aim to systematically assess the accuracy of the ELE approach for realistic molecular models, nor to accurately estimate the computational gain which can be achieved by simulating the effective theory, rather than the full theory. To this purpose, one should perform a systematic analysis based on realistic atomistic models, which include also three-body and four-body potential to account for bonded interactions, along with non-bonded electrostatic forces and solvent induced interactions. The first preliminary results reported here serve as a motivation for such an analysis.

\acknowledgments
This work was motivated by a discussion with A. Szabo.
The path integral of the EST was calculated in collaboration with H. Orland and O. Corradini, who I thankfully acknowledge. 
I thank S.a Beccara, M.Sega, T. Skribic and F. Pederiva for important discussions. P. Faccioli is a member of the Interdisciplinary Laboratory for Computational Sciences (LISC), a joint venture of Trento University and FBK.

 \end{document}